\documentclass
[showpacs,preprintnumbers,nofootinbib,superscriptaddress,onecolumn]{revtex4}%
\usepackage{amsfonts}
\usepackage{amsmath}
\usepackage{amssymb}
\usepackage{graphicx}%
\begin{document}
\title{Remarks about the thermodynamic limit in selfgravitating systems}
\author{L. Velazquez}
\email{luisberis@geo.upr.edu.cu}
\affiliation{Departamento de F\'{\i}sica, Universidad de Pinar del R\'{\i}o, Mart\'{\i}
270, Esq. 27 de Noviembre, Pinar del R\'{\i}o, Cuba.}
\author{F. Guzman}
\affiliation{Departamento de F\'{\i}sica Nuclear, Instituto Superior de Ciencias y
Tecnolog\'{\i}a Nucleares, Carlos III y Luaces, Plaza, La Habana, Cuba.}

\begin{abstract}
The present effort addresses the question about the existence of a
well-defined thermodynamic limit for the astrophysical systems with the
following \textit{power law form}: to tend the number of particles, $N$, the
total energy, $E$, and the characteristic linear dimension of the system,
$L\,$, to infinity, keeping constant $E/N^{\Lambda_{E}}$ and $L/N^{\Lambda
_{L}}$, being $\Lambda_{E}$ and $\Lambda_{L}$ certain scaling exponent
constant. This study is carried out for a system constituted by a non-rotating
fluid under the influence of its own Newtonian gravitational interaction. The
analysis yields that a thermodynamic limit of the above form will only appear
when the local pressure depends on the energy density of fluid as
$p=\gamma\epsilon$, being $\gamma$ certain constant. Therefore, a
thermodynamic limit with a power law form can be only satisfied by a reduced
set of models, such as the selfgravitating gas of fermions and the Antonov
isothermal model.

\end{abstract}
\maketitle

In a series of recent papers, de Vega, Sanchez and Laliena discuss about the
existence of a well-defined thermodynamic limit for the self-gravitating
systems \cite{de Vega,de Vega2,Laliena,Laliena2}. The first two authors
claimed that the thermodynamic limit of a self-gravitating system can be taken
by letting the number of particles, $N$, and the volume, $V$, tend to infinity
keeping the ratio $N/V^{\frac{1}{3}}$ constant. On the other hand, Laliena
exposes a series of reasons which in his viewpoint makes invalid the
consideration of the above thermodynamic limit. The present paper is aimed to
contribute to clarify the question about the existence of a well-defined
thermodynamic limit for self-gravitating systems.

Our analysis starts from the consideration of a mean field approximation for
the estimation of the microcanonical volume $W$ of a non-rotating fluid under
the influence of its own Newtonian gravitational interaction:%

\begin{equation}
W_{MF}\left(  E,N,L\right)  =\int\mathcal{D}\rho\left(  \mathbf{r}\right)
\mathcal{D}\epsilon\left(  \mathbf{r}\right)  \mathcal{D}\phi\left(
\mathbf{r}\right)  ~\delta\left(  E-H[\epsilon,\rho,\phi]\right)
\delta\left(  N-N[\rho]\right)  \delta\left\{  \phi\left(  \mathbf{r}\right)
-\mathcal{G}\left[  \rho,\mathbf{r}\right]  \right\}  \exp\left\{
S[\epsilon,\rho]\right\}  ,
\end{equation}
being $\rho\left(  \mathbf{r}\right)  $ and $\epsilon\left(  \mathbf{r}%
\right)  $, the particles and the internal energy density of the fluid at the
point $\mathbf{r}$, and $\phi\left(  \mathbf{r}\right)  $, the Newtonian
potential at the same point. The functionals:%

\begin{equation}
S\left[  \epsilon,\rho\right]  =\int d^{3}\mathbf{r}~s\left\{  \epsilon\left(
\mathbf{r}\right)  ,\rho\left(  \mathbf{r}\right)  \right\}  ,~H\left[
\epsilon,\rho,\phi\right]  =\int d^{3}\mathbf{r}~\epsilon\left(
\mathbf{r}\right)  +\frac{1}{2}m\rho\left(  \mathbf{r}\right)  \phi\left(
\mathbf{r}\right)  ,~N\left[  \rho\right]  =\int d^{3}\mathbf{r}~\rho\left(
\mathbf{r}\right)  ,
\end{equation}
represent the total entropy, energy and particles number for a given profile
with $\epsilon\left(  \mathbf{r}\right)  $ and $\rho\left(  \mathbf{r}\right)
$; and the functional $\mathcal{G}\left[  \rho;\mathbf{r}\right]  $:%

\begin{equation}
\mathcal{G}\left[  \rho;\mathbf{r}\right]  =-\int\frac{Gm\rho\left(
\mathbf{r}_{1}\right)  d^{3}\mathbf{r}_{1}}{\left\vert \mathbf{r}%
-\mathbf{r}_{1}\right\vert }, \label{green}%
\end{equation}
is the Green solution of the Poisson problem:%

\begin{equation}
\Delta\phi=4\pi Gm\rho.
\end{equation}

This approximation was analyzed in details in ref.\cite{Velazquez}. This
system has been enclosed in a spherical rigid container with characteristic
linear dimension $L$ in order to avoid the long-range divergence of the
Newtonian gravitational potential. On the other hand, the short-range
divergence of this interaction has been also avoided by the consideration of
quantum effects or the nature size of the particles which are compose the
fluid, which are taken in an implicit manner in the entropy density of the
fluid, $s\left\{  \epsilon\left(  \mathbf{r}\right)  ,\rho\left(
\mathbf{r}\right)  \right\}  $.

$W_{MF}\left(  E,N,L\right)  $ can be rewritten by using the Fourier
representation of the delta functions: $\delta\left(  x\right)  =\int
_{-\infty}^{+\infty}\frac{dk}{2\pi}\exp\left(  zx\right)  $, being
$z=\varepsilon+ik$, yielding:%

\begin{equation}
W_{MF}\left(  E,N\right)  \sim\int_{-\infty}^{+\infty}\int_{-\infty}^{+\infty
}\frac{dkd\eta}{\left(  2\pi\right)  ^{2}}\int\mathcal{D}\rho\left(
\mathbf{r}\right)  \mathcal{D}\epsilon\left(  \mathbf{r}\right)
\mathcal{D}\phi\left(  \mathbf{r}\right)  \mathcal{D}h\left(  \mathbf{r}%
\right)  \exp\left\{  \mathcal{L}\left[  \epsilon,\rho,\phi;z,z_{1},J\right]
\right\}  , \label{wm}%
\end{equation}
where $z=\beta+ik$ and $z_{1}=\mu+i\eta$ with $\beta,\eta\in\mathbb{R}$, being
the functional $\mathcal{L}\left[  \epsilon,\rho,\phi;z,z_{1},J\right]  $
defined by:%

\begin{equation}
\mathcal{L}\left[  \epsilon,\rho,\phi;z,z_{1},J\right]  =S\left[
\epsilon,\rho\right]  +z\left(  E-H\left[  \epsilon,\rho,\phi\right]  \right)
+z_{1}\left(  N-N\left[  \rho\right]  \right)  +J\ast\left(  \phi
-\mathcal{G}\left[  \rho\right]  \right)  . \label{Leg}%
\end{equation}
The functional term
\[
J\ast\left(  \phi-\mathcal{G}\left[  \rho\right]  \right)  \equiv\int
d^{3}\mathbf{r}~J\left(  \mathbf{r}\right)  \left\{  \phi\left(
\mathbf{r}\right)  -\mathcal{G}\left[  \rho;\mathbf{r}\right]  \right\}
\]
appears as consequence of the Fourier representation of the delta functional
$\delta\left\{  \phi\left(  \mathbf{r}\right)  -\mathcal{G}\left[
\rho;\mathbf{r}\right]  \right\}  $. Here, $J\left(  \mathbf{r}\right)  $ is a
complex function, $J\left(  \mathbf{r}\right)  =j\left(  \mathbf{r}\right)
+ih\left(  \mathbf{r}\right)  $, with $j\left(  \mathbf{r}\right)
\in\mathbb{R}$.

The direct integration of the functional integrals in the expression
(\ref{wm}) is a formidable task which could be only carried out by using
adequate approximations. The usual methodology which is applied in this
situations is the called \textit{steepest decent method}. The application of
this method is based on the asymptotic behavior of the thermodynamical
variables and potentials in the many particle limit $N\gg1$, which leads to
estimate that the main contribution for the expectation values for the
physical observables in the microcanonical ensemble comes from the most
probably configurations of the system. Thus, the Boltzmann entropy of the
system, $S_{B}=\ln W$, is obtained by using a min-max procedure:%

\begin{equation}
S_{B}\left(  E,N,L\right)  \simeq~\underset{\beta,~\mu,~j}{\min}\left\{
\underset{\epsilon,~\rho,~\phi}{\max}\mathcal{L}\left[  \epsilon,\rho
,\phi;\beta,\mu,j\right]  \right\}  ,
\end{equation}
whose the stationary conditions leads to the structure equations of the
equilibrium configurations. According to the definition of the functional
$\mathcal{L}\left[  \epsilon,\rho,\phi;\beta,\mu,j\right]  $, the equation
(\ref{Leg}), the reader may recognize the Lagrange formalism\ of the Classical Thermodynamics.

The presence of an additive kinetic part in the Hamiltonian of certain system
leads to an exponential growing of the microcanonical volume $W$ with the $N$
increasing, and therefore, the Boltzmann entropy will grow proportional to $N$
in the many particles limit, $S_{B}=\ln W\propto N$. Therefore, let us now
concentrate our attention in the analysis of the \textit{N}-behavior of the
thermodynamical variables with the growing of the number of particles. 

The usual thermodynamic limit for the extensive systems:%

\begin{equation}
N\rightarrow\infty,\text{ keeping constant }\frac{E}{N}\text{ and }\frac{N}%
{V},
\end{equation}
where $V$ is the volume of the system, is directly related with the
\textit{extensive properties} of these systems when the thermodynamical
variables of the system are scaled by some scaling parameter $\alpha$ as follows:%

\begin{equation}
\left.
\begin{array}
[c]{c}%
N\rightarrow N\left(  \alpha\right)  =\alpha N_{,}\\
E\rightarrow E\left(  \alpha\right)  =\alpha E,\\
V\rightarrow V\left(  \alpha\right)  =\alpha V,
\end{array}
\right\}  \Rightarrow W\rightarrow W\left(  \alpha\right)  =\exp\left(
\alpha\ln W\right)  . \label{extensive}%
\end{equation}
In analogy with the extensive properties of the traditional systems, we will
analyze the necessary conditions for the existence of the following
\textit{power law self-similarity }scaling behavior of the microcanonical
variables $E$, $N$ and $L$ for the selfgravitating systems:%

\begin{equation}
\left.
\begin{array}
[c]{c}%
N\rightarrow N\left(  \alpha\right)  =\alpha N_{,}\\
E\rightarrow E\left(  \alpha\right)  =\alpha^{\Lambda_{E}}E,\\
L\rightarrow L\left(  \alpha\right)  =\alpha^{\Lambda_{L}}L,
\end{array}
\right\}  \Rightarrow W\rightarrow W\left(  \alpha\right)  =\exp\left(
\alpha\ln W\right)  , \label{self}%
\end{equation}
where $\Lambda_{E}$ and $\Lambda_{L}$ are certain constant scaling exponent
which lead to an \textit{extensive }character of the Boltzmann entropy. This
kind of self-similarity behavior is directly related with a thermodynamic
limit with a power law form:%

\begin{equation}
N\rightarrow\infty\text{, keeping constant }\frac{E}{N^{\Lambda_{E}}}\text{
and }\frac{L}{N^{\Lambda_{L}}}. \label{THL}%
\end{equation}

The existence of this kind of self-similarity condition allows a considerable
simplification of the thermodynamical description: the study can be performed
by setting $N=1$ and considering the \textit{N}-dependence in the scaling laws
by taking $\alpha=N$. This scaling behavior is very useful in numerical
experiments, since it allows us to extend the results of this kind of study on
a finite system to much bigger systems. Contrary, the nontrivial
\textit{N-}dependent behavior of the thermodynamical variables and potentials
leads to a complication of any kind of study.

In order to satisfy this scaling behavior for the global variables $E$, $N$,
$L$ and the Boltzmann entropy $S_{B}$, the local functions $\epsilon\left(
\mathbf{r}\right)  $, $\rho\left(  \mathbf{r}\right)  $, $\phi\left(
\mathbf{r}\right)  $ and $s\left(  \epsilon,\rho;\phi\right)  $ should be
scaled as follows:%

\begin{equation}
\left.  \rho\rightarrow\rho\left(  \alpha\right)  =\alpha^{\Lambda_{\rho}}%
\rho\right\}  \Rightarrow\left\{
\begin{array}
[c]{c}%
\phi\rightarrow\phi\left(  \alpha\right)  =\alpha^{\Lambda_{\phi}}\phi\\
\epsilon\rightarrow\epsilon\left(  \alpha\right)  =\alpha^{\Lambda_{e}%
}\epsilon\\
s\rightarrow s\left(  \alpha\right)  =\alpha^{\Lambda_{S}}s
\end{array}
\right\}  .
\end{equation}

Since the characteristic particles density behaves as $\rho_{c}\sim N/L^{3}$,
the scaling exponent for the particles density is $\Lambda_{\rho}%
=1-3\Lambda_{L}$. From the expression of the Newtonian potential (\ref{green})
is derived that its characteristic unit is $\phi_{c}\sim\rho_{c}L^{2}$, and
therefore, $\Lambda_{\phi}=1-\Lambda_{L}$. The energy scaling exponent is
equal to the scaling exponent of the total gravitational potential energy, so
that, $\Lambda_{E}=2-\Lambda_{L}$. The other scaling exponents are obtained by
using identical reasonings. All these scaling exponents depend on the scaling
exponent $\Lambda_{L}$ as follows:%

\begin{equation}%
\begin{array}
[c]{c}%
\Lambda_{\rho}=1-3\Lambda_{L}=\Lambda_{S},~\Lambda_{\phi}=1-\Lambda_{L},\\
\Lambda_{e}=2-4\Lambda_{L},~\Lambda_{E}=2-\Lambda_{L}.
\end{array}
~
\end{equation}

In order to satisfy these scaling laws is also \textit{necessary} that the
entropy density exhibits to the following scaling behavior:%

\begin{equation}
s\left(  \alpha^{\Lambda_{e}}\epsilon,\alpha^{\Lambda_{\rho}}\rho\right)
=\alpha^{\Lambda_{\rho}}s\left(  \epsilon,\rho\right)  ,
\end{equation}
This scaling property is satisfy if $s\left(  \epsilon,\rho\right)  $ obeys to
the following functional form:%

\begin{equation}
s\left(  \epsilon,\rho\right)  =\rho F\left(  \epsilon/\rho^{\eta}\right)  ,
\label{ent ff}%
\end{equation}
where $\eta=\Lambda_{e}/\Lambda_{\rho}$, being $F\left(  x\right)  $ an
arbitrary function. This functional form leads to a simple relation between
the local pressure $p$ and the internal energy density $\epsilon$ of the
fluid. By taking into account the functional form (\ref{ent ff}) and using the relations%

\begin{equation}
\beta=\frac{\partial}{\partial\epsilon}s\left(  \epsilon,\rho\right)  ,~\beta
p=s\left(  \epsilon,\rho\right)  -\epsilon\frac{\partial}{\partial\epsilon
}s\left(  \epsilon,\rho\right)  -\rho\frac{\partial}{\partial\rho}s\left(
\epsilon,\rho\right)  ,
\end{equation}
it is straightforward derived the relation:%

\begin{equation}
p=\gamma\epsilon, \label{gam}%
\end{equation}
where $\gamma=\eta-1$. There are some well-known Hamiltonian systems which
satisfy this kind of relation, as example, the system of nonrelativistic or
ultrarelativistic noninteracting particles, without matter if they obey to the
Boltzmann, Fermi-Dirac or Bose-Einstein Statistics. The scaling parameter
$\Lambda_{L}$ and $\Lambda_{E}$ are obtained from the parameter $\gamma$ as follows:%

\begin{equation}
\Lambda_{L}=\frac{\gamma-1}{3\gamma-1},~\,\Lambda_{E}=\frac{5\gamma-1}%
{3\gamma-1}.
\end{equation}

This result evidences that the power laws form for the self-similarity
conditions (\ref{self}) can be only satisfied by a reduced group of models
whose microscopic picture obeys to the relation\ (\ref{gam}). Since
$\gamma=\frac{2}{3}$ for the ideal gas of particles, the scaling exponents for
the Antonov problem \cite{antonov} and the selfgravitating fermions model
\cite{thir2,bilic} are given by $\Lambda_{E}=\frac{7}{3}$ and $\Lambda
_{L}=-\frac{1}{3}$, and therefore, they obey to the following thermodynamic limit:%

\begin{equation}
N\rightarrow\infty\text{, keeping constant }\frac{E}{N^{\frac{7}{3}}}\text{
and }LN^{\frac{1}{3}}\text{.} \label{thermodynamic limit}%
\end{equation}
This thermodynamic limit was established in the ref.\cite{chava} for the
self-gravitating nonrelativistic fermions by using other reasonings. Since the
selfgravitating relativistic gas and classic hard sphere model
\cite{oroson,stahl} do not obey to the relation (\ref{gam}), \textit{they do
not posses a scaling behavior with a\ power law form} (\ref{self}).

This result clarifies a question about the discussion of the adequate
thermodynamic limit for selfgravitating systems. All above selfgravitating
models converge in the low density limit in the selfgravitating gas, and
therefore, they exhibit a \textit{N}-growing of the energy with\ the $7/3$
power law. However, they diverge in regard to the high density limit since
them used different regularization procedure for the short-range divergence of
the Newtonian potential. Generally speaking, such power law thermodynamic
limit (\ref{THL}) does not exist in selfgravitating systems for the whole
values of the microcanonical variables. Even the selfgravitating gas of
fermions exhibits this behavior only in the nonrelativistic limit since the
relation (\ref{ent ff}) disappears when the relativistic effects are taken
into account for a massive enough selfgravitating system.

\end{document}